\documentclass[a4paper]{article}
\RequirePackage[english]{babel}
\RequirePackage[latin1]{inputenc}
\RequirePackage[T1]{fontenc}
\RequirePackage{mathrsfs}
\RequirePackage{amsmath}
\RequirePackage{amssymb}
\RequirePackage{amsbsy}
\RequirePackage{bm}
\begin{document}
\title{\bf{On a purely geometric approach to the\\ Dirac matter field and its\\ 
quantum properties}}
\author{Luca Fabbri\\ 
\footnotesize INFN, Sez. di Bologna \& Dipartimento di Fisica, Universit\`{a} di Bologna, 
Bologna, Italy}
\date{}
\maketitle
\begin{abstract}
We consider the most general axial torsion completion of gravity with electrodynamics for $\frac{1}{2}$-spin spinors in an $8$-dimensional representation of the Dirac matter field: this theory will allow to define antimatter as matter with all quantum numbers reversed, where also the sign of the mass beside that of the charge is inverted: we shall see that matter and antimatter solutions of the Dirac field equations coincide with the known ones with respect to all observables, that despite the inversion of the sign of the mass term only positive-mass states are present and only positive-energy densities are given; the present and the common approach will be compared, and some experimental implications will be discussed.
\end{abstract}
\section*{Introduction}
If we were to ask mathematicians or philosophers what would be the single most fundamental argument in the geometry of absolute differential calculus for the presence of torsion, they would reverse the question by asking what would be the arguments \emph{not} to have torsion instead: in fact if we want to develop a geometry for tensors and their derivatives, we need a connection which, in its most general form, has torsion already; although there is no \emph{a priori} reason to have torsion removed, nevertheless there have been attempts to invoke some principle to deduce that torsion must vanish, and indeed for an unambiguous implementation of the principles of equivalence and causality a sufficient condition is to have torsion equal to zero \cite{m-t-w}: nevertheless, if we want to be in the most general case with a sufficient and necessary condition then torsion need only be completely antisymmetric \cite{ha,xy,a-l,m-l,f/1a,f/1b}. If we were to ask theoreticians what would be the most important consequence for physics of the presence of torsion, they would state that torsion enters beside curvature and gauge fields to complete the set of geometrical fields, on the one hand, much like the spin is present alongside the energy and the current as the complete set of matter conserved quantities, on the other hand, so that they may be fully coupled, according to the prescription to have respectively torsion-spin, curvature-energy and gauge-current dynamical field equations, as discussed in \cite{h-h-k-n,h-o,Capozziello:2011et}; because the only field for which the principle of equivalence and causality are respected without any restriction is the 
$\frac{1}{2}$-spin spinor, which has a completely antisymmetric spin, then both torsion and the spin are completely antisymmetric, and the torsion-spin coupling field equations still have their validity \cite{f/2a,f/2b}. If we were to ask phenomenologists what would be the effects torsion would have for observations, they would reply that evidence should be sought in the torsionally-induced spin-contact interactions among spinors, such as for instance a dynamical form of the principle of exclusion for fermions described by Dirac matter fields \cite{f/3}. This makes clear the importance of the torsionally-induced non-linear potentials for the dynamics of the matter fields. For long the only weakest point of torsion was that its effects were believed to be relevant only at the Planck scale, until recent developments in which more general dynamics for torsion were considered opened the way toward a solution \cite{Baekler:2011jt,k}: in such generalization, the torsionally-induced spin-contact interactions among spinors have a constant that is yet to be determined, and so potentially relevant at larger scales \cite{Fabbri:2011kq}. As a consequence of the fact that torsion is algebraically related to the spin, which is algebraically given in terms of bilinear spinors, torsion vanishes where there is no spin, that is out of spinorial distributions, and thus, although the effects for the fermions can be amplified within the Dirac matter field distribution, they are absent in vacuum, compatibly with all the experimental limits that are presently known.

As it stands, this theory appears to be the single generalization that comes from first principles and which is compatible with all observational evidence, able to provide non-linear interactions among fermions within the Dirac matter field equation, identical to those already investigated in renowned theories such as Nambu-Jona--Lasinio model, and in this paper, we use this theory to suggest an alternative definition of the matter/antimatter duality.
\section{Kinematical Symmetries}
The notation we will employ is that of \cite{Fabbri:2011kq}, now recalling some definition.

In everything that follows, the Cartan-Riemann geometry is taken in its most general form, with metric tensors $g_{\alpha\sigma}$ and $g^{\alpha\sigma}$ symmetric and one the inverse of the other, a connection $\Gamma^{\alpha}_{\mu\nu}$ defining a covariant derivative $D_{\mu}$ for which metric-compatibility $Dg\!=\!0$ holds and such that the Cartan torsion tensor
\begin{eqnarray}
&Q^{\alpha}_{\phantom{\alpha}\mu\nu}=\Gamma^{\alpha}_{\mu\nu}-\Gamma^{\alpha}_{\nu\mu}
\label{Cartan}
\end{eqnarray}
is taken to be completely antisymmetric $Q_{[\alpha\mu\rho]}=6Q_{\alpha\mu\rho}$ identically: the metric-compatibility condition and complete antisymmetry of torsion encode the fact that there exists a unique symmetric part of the connection that can be vanished while the metric can be flattened in the same neighborhood of the very same coordinate system, and then this connection can be decomposed according to
\begin{eqnarray}
&\Gamma^{\mu}_{\sigma\pi}
=\frac{1}{2}Q^{\mu}_{\phantom{\mu}\sigma\pi}
+\frac{1}{2}g^{\mu\rho}\left(\partial_{\pi}g_{\sigma\rho}
+\partial_{\sigma}g_{\pi\rho}-\partial_{\rho}g_{\sigma\pi}\right)
\label{connection}
\end{eqnarray}
which holds in general; we can define the Riemann curvature tensor
\begin{eqnarray}
&G^{\mu}_{\phantom{\mu}\rho\sigma\pi}=\partial_{\sigma}\Gamma^{\mu}_{\rho\pi}
-\partial_{\pi}\Gamma^{\mu}_{\rho\sigma}
+\Gamma^{\mu}_{\lambda\sigma}\Gamma^{\lambda}_{\rho\pi}
-\Gamma^{\mu}_{\lambda\pi}\Gamma^{\lambda}_{\rho\sigma}
\label{Riemann}
\end{eqnarray}
antisymmetric in the first and second couple of indices, with one independent contraction given by $\!G^{\alpha}_{\phantom{\alpha}\rho\alpha\sigma}\!\!=\!G_{\rho\sigma}$ with $\!G_{\rho\sigma}g^{\rho\sigma}\!\!=\!G$ named Ricci curvature tensor and scalar, and for which we have the decomposition given by the expression
\begin{eqnarray}
&G^{\mu}_{\phantom{\mu}\rho\sigma\pi}\!=\!R^{\mu}_{\phantom{\mu}\rho\sigma\pi}
\!+\!\frac{1}{2}(\nabla_{\sigma}Q^{\mu}_{\phantom{\mu}\rho\pi}
-\nabla_{\pi}Q^{\mu}_{\phantom{\mu}\rho\sigma})
\!+\!\frac{1}{4}(Q^{\mu}_{\phantom{\mu}\lambda\sigma}Q^{\lambda}_{\phantom{\lambda}\rho\pi}
-Q^{\mu}_{\phantom{\mu}\lambda\pi}Q^{\lambda}_{\phantom{\lambda}\rho\sigma})
\label{decomposition}
\end{eqnarray}
given in terms of the Levi-Civita metric covariant derivative $\nabla_{\nu}$ and the Riemann metric curvature tensor $R^{\mu}_{\phantom{\mu}\rho\sigma\pi}$ whose contraction is correspondingly given according to $R^{\alpha}_{\phantom{\alpha}\rho\alpha\sigma}\!\!=\!R_{\rho\sigma}$ with $R_{\rho\sigma}g^{\rho\sigma}\!\!=\!R$ called Ricci metric curvature tensor and scalar as they are commonly called in the literature. An equivalent formalism is defined by the pair of dual bases of tetrads $\xi^{a}_{\sigma}$ and $\xi_{a}^{\sigma}$ verifying orthonormality conditions $\xi_{a}^{\sigma}\xi_{b}^{\nu}g_{\sigma\nu}\!=\!\eta_{ab}$ and $\xi^{a}_{\sigma}\xi^{b}_{\nu}g^{\sigma\nu}\!=\!\eta^{ab}$ in terms of the Minkowskian matrices, and a spin-connection $\Gamma^{i}_{j\mu}$ defining the covariant derivative $D_{\mu}$ such that it gives $D\xi\!=\!0$ and $D\eta\!=\!0$ and in this formalism it is not possible to define a torsion tensor, although (\ref{Cartan}) can be written according to
\begin{eqnarray}
&-Q^{a}_{\phantom{a}\mu\nu}=\partial_{\mu}\xi^{a}_{\nu}-\partial_{\nu}\xi^{a}_{\mu}
+\Gamma^{a}_{j\mu}\xi^{j}_{\nu}-\Gamma^{a}_{j\nu}\xi^{j}_{\mu}
\label{Cartangauge}
\end{eqnarray}
as it might have been expected: these conditions imply that (\ref{connection}) is given by
\begin{eqnarray}
&\Gamma^{b}_{\phantom{b}j\mu}=
\xi^{\alpha}_{j}\xi_{\rho}^{b}\left(\Gamma^{\rho}_{\phantom{\rho}\alpha\mu}
+\xi_{\alpha}^{k}\partial_{\mu}\xi^{\rho}_{k}\right)
\label{spin-connection}
\end{eqnarray}
and it is antisymmetric in the two Lorentz indices; the curvature is
\begin{eqnarray}
&G^{a}_{\phantom{a}b\sigma\pi}
=\partial_{\sigma}\Gamma^{a}_{b\pi}-\partial_{\pi}\Gamma^{a}_{b\sigma}
+\Gamma^{a}_{j\sigma}\Gamma^{j}_{b\pi}-\Gamma^{a}_{j\pi}\Gamma^{j}_{b\sigma}
\label{Riemanngauge}
\end{eqnarray}
antisymmetric in both the coordinate and the Lorentz indices and writable as
\begin{eqnarray}
&G^{a}_{\phantom{a}b\sigma\pi}=G^{\mu}_{\phantom{\mu}\rho\sigma\pi}\xi^{\rho}_{b}\xi^{a}_{\mu}
\label{correlation}
\end{eqnarray}
in terms of the previous expression (\ref{Riemann}) of the Riemann tensor. The former geometrical setting defined in terms of Greek indices was covariant under the most general coordinate transformation and its translation into the equivalent formalism defined in terms of Latin indices has the advantage that the covariance is translated into covariance under special Lorentz transformations, which can be written explicitly and therefore they can also be written in terms of other representations, such as the complex one. Of course, when complex representations are considered, complex fields must be introduced, and in an analogous way we may also define the geometry of complex fields, where the introduction of the gauge-connection $A_{\mu}$ defines the gauge-covariant derivative $D_{\mu}$ extending differential properties to complex fields; its curvature is given by the expression
\begin{eqnarray}
&F_{\mu\nu}=\partial_{\mu}A_{\nu}-\partial_{\nu}A_{\mu}
\label{Maxwellgauge}
\end{eqnarray}
antisymmetric in the two indices, called Maxwell tensor. Next step will be to consider the Lorentz transformation and explicitly construct its complex representation, which will have to be taken together with the gauge transformation, to form the most general transformation we may have for matter fields.

To build this representation, we start by taking its simplest form given by the irreducible representation written in terms of the $\frac{1}{2}$-spin $2$-dimensional sigma matrices $\vec{\boldsymbol{\sigma}}$ known as Pauli matrices: we may construct the most general irreducible $\frac{1}{2}$-spin $2$-dimensional transformation by taking the infinitesimal generators as given by a basis for such a space, and since such a basis is precisely given by the $\vec{\boldsymbol{\sigma}}$ and the identity matrix then we have that the most general transformation is given by $\mathrm{exp}[(\vec{\varphi}\!+\!i\vec{\theta})\!\cdot\!
\vec{\frac{\boldsymbol{\sigma}}{2}}\!+\!(\beta\!+\!i\alpha)\mathbb{I}]$ with all parameters being real sets of functions; however, we notice that the $\beta$ parameter will describe a scaling that would give rise to a conformal structure which should not be present in our description, and so we require such function to vanish: the most general non-conformal transformation is given by $\mathrm{exp}[(\vec{\varphi}\!+\!i\vec{\theta})\!\cdot\! \vec{\frac{\boldsymbol{\sigma}}{2}}
\!+\!i\alpha\mathbb{I}]$ consisting of the Lorentz complex representation with the additional term proportional to the identity matrix describing the gauge transformation. In the following we are going to drop for convenience the presence of the identity matrix and we will introduce the label $q$ called charge: thus we may write the most general non-conformal transformation according to $\mathrm{exp}[(\vec{\varphi}\!+\!i\vec{\theta})\!\cdot\!\vec{\frac{\boldsymbol{\sigma}}{2}}
\!+\!i\alpha q]$ where all parameters are real functions. We notice however that although this is the most general non-conformal transformation, nevertheless there is still a problem we have to address. This problem consists in the fact that such a transformation is not uniquely defined: as a quick inventory would clearly show, transformations given by $\mathrm{exp}[(\vec{\varphi}
\!+\!i\vec{\theta})\!\cdot\!\vec{\frac{\boldsymbol{\sigma}}{2}}\!+\!i\alpha q]$ or 
$\mathrm{exp}[(-\vec{\varphi}\!+\!i\vec{\theta})\!\cdot\!\vec{\frac{\boldsymbol{\sigma}}{2}}
\!+\!i\alpha q]$ but also the transformations given by $\mathrm{exp}[(\vec{\varphi}
\!+\!i\vec{\theta})\!\cdot\!\vec{\frac{\boldsymbol{\sigma}}{2}}\!-\!i\alpha q]$ or 
$\mathrm{exp}[(-\vec{\varphi}\!+\!i\vec{\theta})\!\cdot\!\vec{\frac{\boldsymbol{\sigma}}{2}}
\!-\!i\alpha q]$ are all possible, and as it is also clear these four transformations exhaust all the relative sign combinations: from a physical point of view, this $4$-fold multiplicity comes from the fact that a particle with a certain spin can be described in terms of two opposite helicity states, and in each case we have that both positive and negative charges are possible. The transformations corresponding to left-handed and right-handed helicities are two irreducible $2$-dimensional chiral components that can be merged into a reducible $4$-dimensional chiral representation as
\begin{eqnarray}
\vec{\boldsymbol{\gamma}}_{\mathrm{ch}}
=\left(\begin{array}{cc}
\boldsymbol{0} & \vec{\boldsymbol{\sigma}}\\
-\vec{\boldsymbol{\sigma}} & \boldsymbol{0}
\end{array}\right)\ \ \ \ \ \ \ \ 
\boldsymbol{\gamma}^{0}_{\mathrm{ch}}
=\left(\begin{array}{cc}
\boldsymbol{0} & \boldsymbol{\mathbb{I}}\\
\boldsymbol{\mathbb{I}} & \boldsymbol{0}
\end{array}\right)
\end{eqnarray}
or the unitarily equivalent standard representation
\begin{eqnarray}
\vec{\boldsymbol{\gamma}}_{\mathrm{st}}
=\left(\begin{array}{cc}
\boldsymbol{0} & -\vec{\boldsymbol{\sigma}}\\
\vec{\boldsymbol{\sigma}} & \boldsymbol{0}
\end{array}\right)\ \ \ \ \ \ \ \ 
\boldsymbol{\gamma}^{0}_{\mathrm{st}}
=\left(\begin{array}{cc}
\boldsymbol{\mathbb{I}} & \boldsymbol{0}\\
\boldsymbol{0} & -\boldsymbol{\mathbb{I}}
\end{array}\right)
\end{eqnarray}
both defining matrices $\frac{1}{4}[\boldsymbol{\gamma}_{i},\boldsymbol{\gamma}_{j}]\!=\!\boldsymbol{\sigma}_{ij}$ so that $\{\boldsymbol{\gamma}_{i},\boldsymbol{\sigma}_{jk}\}\!=\!i\varepsilon_{ijkq}\boldsymbol{\gamma}\boldsymbol{\gamma}^{q}$ in terms of which we have that the pair of opposite-helicity Lorentz representations can be written as $\mathrm{exp}[\frac{1}{2}\theta^{ij}\boldsymbol{\sigma}_{ij}+i\alpha q]$ or 
$\mathrm{exp}[\frac{1}{2}\theta^{ij}\boldsymbol{\sigma}_{ij}-i\alpha q]$ altogether; these correspond to the transformation of two $4$-dimensional components of opposite charge which can therefore be merged into an $8$-dimensional chiral representation
\begin{eqnarray}
\nonumber
&\vec{\boldsymbol{\Gamma}}_{\mathrm{ch}}
=\left(\begin{array}{cc}
\vec{\boldsymbol{\gamma}}_{\mathrm{ch}} & \boldsymbol{0}\\
\boldsymbol{0} & \vec{\boldsymbol{\gamma}}_{\mathrm{ch}}
\end{array}\right)\ \ \ \ \ \ \ \ 
\boldsymbol{\Gamma}^{0}_{\mathrm{ch}}
=\left(\begin{array}{cc}
\boldsymbol{\gamma}^{0}_{\mathrm{ch}} & \boldsymbol{0}\\
\boldsymbol{0} & \boldsymbol{\gamma}^{0}_{\mathrm{ch}}
\end{array}\right)\\
&\boldsymbol{K}_{\mathrm{ch}}
=\left(\begin{array}{cc}
\boldsymbol{\mathbb{I}} & \boldsymbol{0}\\
\boldsymbol{0} & -\boldsymbol{\mathbb{I}}
\end{array}\right)
\end{eqnarray}
or a unitarily equivalent first type of standard representation as
\begin{eqnarray}
\nonumber
&\vec{\boldsymbol{\Gamma}}_{\mathrm{st1}}
=\left(\begin{array}{cc}
\vec{\boldsymbol{\gamma}}_{\mathrm{st}} & \boldsymbol{0}\\
\boldsymbol{0} & \vec{\boldsymbol{\gamma}}_{\mathrm{st}}
\end{array}\right)\ \ \ \ \ \ \ \ 
\boldsymbol{\Gamma}^{0}_{\mathrm{st1}}
=\left(\begin{array}{cc}
\boldsymbol{\gamma}^{0}_{\mathrm{st}} & \boldsymbol{0}\\
\boldsymbol{0} & \boldsymbol{\gamma}^{0}_{\mathrm{st}}
\end{array}\right)\\
&\boldsymbol{K}_{\mathrm{st1}}
=\left(\begin{array}{cc}
\boldsymbol{\mathbb{I}} & \boldsymbol{0}\\
\boldsymbol{0} & -\boldsymbol{\mathbb{I}}
\end{array}\right)
\end{eqnarray}
or yet another unitarily equivalent second type of standard representation
\begin{eqnarray}
\nonumber
&\vec{\boldsymbol{\Gamma}}_{\mathrm{st2}}
=\left(\begin{array}{cc}
\boldsymbol{0} & \vec{\boldsymbol{\gamma}}_{\mathrm{st}}\\
\vec{\boldsymbol{\gamma}}_{\mathrm{st}} & \boldsymbol{0}
\end{array}\right)\ \ \ \ \ \ \ \ 
\boldsymbol{\Gamma}^{0}_{\mathrm{st2}}
=\left(\begin{array}{cc}
\boldsymbol{\gamma}^{0}_{\mathrm{st}} & \boldsymbol{0}\\
\boldsymbol{0} & \boldsymbol{\gamma}^{0}_{\mathrm{st}}
\end{array}\right)\\
&\boldsymbol{K}_{\mathrm{st2}}
=\left(\begin{array}{cc}
\boldsymbol{\gamma}^{0}_{\mathrm{st}} & \boldsymbol{0}\\
\boldsymbol{0} & -\boldsymbol{\gamma}^{0}_{\mathrm{st}}
\end{array}\right)
\end{eqnarray}
all defining $[\boldsymbol{\Gamma}_{i},\boldsymbol{\Gamma}_{j}]\!=\!4\boldsymbol{\Sigma}_{ij}$ so that $\{\boldsymbol{\Gamma}_{i},\boldsymbol{\Sigma}_{jk}\}\!=\!i\varepsilon_{ijkq} \boldsymbol{\Gamma}\boldsymbol{\Gamma}^{q}$ and thus the complete Lorentz complex representation is $\mathrm{exp}[\frac{1}{2}\theta^{ij}\boldsymbol{\Sigma}_{ij}+iq\alpha\boldsymbol{K}]$ usually called spinorial representation, and this is the most general non-conformal complex Lorentz transformation we will employ in the following. We also introduce the matrix
\begin{eqnarray}
\boldsymbol{M}
=\left(\begin{array}{cc}
\boldsymbol{0} & \boldsymbol{\mathbb{I}}\\
\boldsymbol{\mathbb{I}} & \boldsymbol{0}
\end{array}\right)
\end{eqnarray}
with the same form in any of the above representations, and which will become useful in the following of the paper as we will see. We will define a spinor field as what transforms according to such a spinorial transformation law, and again it is possible to introduce the spinor-connection $\boldsymbol{A}_{\mu}$ through which we define the spinor-covariant derivative $\boldsymbol{D}_{\mu}$ containing the information about the dynamics of the spinor fields: the spinorial constancy of the matrices $\boldsymbol{\Gamma}_{j}$ is implemented automatically, and thus the spinor-connection $\boldsymbol{A}_{\mu}$ is decomposed according to
\begin{eqnarray}
&\boldsymbol{A}_{\mu}=\frac{1}{2}\Gamma^{ab}_{\phantom{ab}\mu}\boldsymbol{\Sigma}_{ab}
+iqA_{\mu}\boldsymbol{K}
\label{spinor-connection}
\end{eqnarray}
in terms of the complex-valued spin-connection plus an abelian field which we may finally identify with the Maxwell gauge-connection; the curvature
\begin{eqnarray}
&\boldsymbol{F}_{\sigma\pi}
=\partial_{\sigma}\boldsymbol{A}_{\pi}-\partial_{\pi}\boldsymbol{A}_{\sigma}
+[\boldsymbol{A}_{\sigma},\boldsymbol{A}_{\pi}]
\label{RiemannMaxwellgauge}
\end{eqnarray}
is a tensorial spinor antisymmetric in the tensorial indices writable as
\begin{eqnarray}
&\boldsymbol{F}_{\sigma\pi}=\frac{1}{2}G^{ab}_{\phantom{ab}\sigma\pi}\boldsymbol{\Sigma}_{ab}
+iqF_{\sigma\pi}\boldsymbol{K}
\label{combination}
\end{eqnarray}
as a combination of Riemann and Maxwell tensors. Such a compact form makes it clear that the torsion and curvature tensors, together with the gauge strength, can be incorporated within a single scheme in a very natural way.
\section{Dynamical Equations}
In the section above we recalled some convention about the kinematic setting, and next we will introduce the dynamical action following \cite{Fabbri:2011kq} for the gravitational background but discussing some generalization for the material content.

To implement the dynamics, we require a link between the geometric fields and the material quantities: this is done by considering the least-order derivative lagrangian in its most general form, with quadratic completely antisymmetric torsion and linear curvature, plus the usual term for gauge fields, yielding the field equations for the completely antisymmetric torsion coupled to the spin
\begin{eqnarray}
&Q^{\rho\mu\nu}=-aS^{\rho\mu\nu}
\end{eqnarray}
the field equations for the non-symmetric curvature coupled to the energy
\begin{eqnarray}
\nonumber
&\left(\frac{8\pi k}{a}\!-\!\frac{1}{2}\right)\!\left(\frac{1}{4}\delta^{\mu}_{\nu}Q^{2}
\!-\!\frac{1}{2}Q^{\mu\alpha\sigma}Q_{\nu\alpha\sigma}
\!+\!D_{\rho}Q^{\rho\mu}_{\phantom{\rho\mu}\nu}\right)
\!+\!\left(G^{\mu}_{\phantom{\mu}\nu}-\frac{1}{2}\delta^{\mu}_{\nu}G
-\lambda\delta^{\mu}_{\nu}\right)+\\
&+8\pi k\left(F^{\rho\mu}F_{\rho\nu}-\frac{1}{4}\delta^{\mu}_{\nu}F^{2}\right)
=8\pi kT^{\mu}_{\phantom{\mu}\nu}
\end{eqnarray}
together with the field equations for the gauge fields coupled to the current
\begin{eqnarray}
&\frac{1}{2}F_{\alpha\mu}Q^{\alpha\mu\rho}+D_{\sigma}F^{\sigma\rho}=J^{\rho}
\end{eqnarray}
where the torsional coupling constant $a$ is not fixed, $k$ is the gravitational constant and the parameter $\lambda$ is called the cosmological constant, and in which we have that the conserved quantities are given by the completely antisymmetric spin density $S^{\rho\mu\nu}$ and the non-symmetric energy density $T^{\mu\nu}$ together with the gauge current $J^{\rho}$ as usual; by considering only the geometrical identities known as Jacobi-Bianchi identities, it is possible to prove that these conserved quantities have to verify the conservation laws given according to 
\begin{eqnarray}
&D_{\rho}S^{\rho\mu\nu}+\frac{1}{2}\left(T^{\mu\nu}-T^{\nu\mu}\right)\equiv0
\label{conservationspin}
\end{eqnarray}
and
\begin{eqnarray}
&D_{\mu}T^{\mu\nu}
+T_{\rho\beta}Q^{\rho\beta\nu}-S_{\mu\rho\beta}G^{\mu\rho\beta\nu}+J_{\rho}F^{\rho\nu}\equiv0
\label{conservationenergy}
\end{eqnarray}
together with
\begin{eqnarray}
&D_{\rho}J^{\rho}=0
\label{conservationcurrent}
\end{eqnarray}
verified once matter fields satisfy suitable matter field equations. In the following we are going to look for the matter fields that will satisfy such constraints.

As we have done in the previous section, we have seen that there are four independent ways in which a spinor field may transform, therefore there are four independent spinor fields, and thus we have to expect that there be four independent spinorial field equations in total: of these four independent spinor fields, there are two opposite-helicity semi-spinors for each of the two opposite-charge spinors, so that we will indicate the left-handed and right-handed chiral projections of corresponding negative and positive sign of the charge according to the notation $\psi_{L}^{N}$, $\psi_{R}^{N}$ and $\psi_{L}^{P}$, $\psi_{R}^{P}$ respectively; as it is easy to check by having a look at their transformations, the most general least-order derivative field equations are $i\hbar \boldsymbol{\gamma}^{\mu} \boldsymbol{D}_{\mu}\psi_{L}^{N}\!=\!\mu \psi_{R}^{N}$ and $i\hbar\boldsymbol{\gamma}^{\mu} \boldsymbol{D}_{\mu}\psi_{R}^{N}\!=\!\nu \psi_{L}^{N}$ as well as the complementary $i\hbar \boldsymbol{\gamma}^{\mu} 
\boldsymbol{D}_{\mu}\psi_{L}^{P}\!=\!\zeta \psi_{R}^{P}$ and $i\hbar\boldsymbol{\gamma}^{\mu} \boldsymbol{D}_{\mu}\psi_{R}^{P}\!=\!\eta \psi_{L}^{P}$ as clear. In the Wigner classification of particles given by irreducible representations of the full Poincar\'{e} plus gauge group, a particle is identified once we know its spin and mass beside all charges, that is in the definition of a particle both spin and all its helicity components together with mass and all charges have to be considered as fundamental quantum numbers; now if the heuristic definition of antimatter as matter with all quantum numbers reversed is to be taken in complete generality, the passage from matter to antimatter is to be accomplished by reverting the sign of the helicity and mass as well as the charge: because this is what general arguments dictate, in this paper we will follow such a prescription without paying attention to contradictions that may seem to arise, hoping that whatever problem we may face will be only apparent, and eventually overcome. Now back to the problem of finding the field equations, we may consider the equations we wrote above, taking their squared in order to built a Klein-Gordon equation for each component of the spinors: the requirement that both helicity states of the spinor describe a field of mass $m$ while both helicity states of the antispinor describe a field of mass $-m$ gives $\mu\!=\!\nu\!=\!m$ and $\zeta\!=\!\eta\!=\!-m$ as the most general conditions, giving $i\hbar \boldsymbol{\gamma}^{\mu} \boldsymbol{D}_{\mu}\psi_{L}^{N}\!-\!m\psi_{R}^{N}\!=\!0$ and $i\hbar\boldsymbol{\gamma}^{\mu} \boldsymbol{D}_{\mu}\psi_{R}^{N}\!-\!m\psi_{L}^{N}\!=\!0$ and also the complementary $i\hbar \boldsymbol{\gamma}^{\mu} \boldsymbol{D}_{\mu}\psi_{L}^{P}\!+\!m\psi_{R}^{P}\!=\!0$ and $i\hbar\boldsymbol{\gamma}^{\mu} \boldsymbol{D}_{\mu}\psi_{R}^{P}\!+\!m\psi_{L}^{P}\!=\!0$ as the set of four possible field equations we have been looking for. We notice that there are two discrete symmetries shuffling these four fields, given according to the transformation laws $\psi_{R}^{N}\!\leftrightarrow\!\psi_{L}^{N}$ and $\psi_{R}^{P}\!\leftrightarrow\!\psi_{L}^{P}$ swapping the two chiral projections and also $\psi^{P}\!\leftrightarrow\!\psi^{N}$ with 
$m\!\rightarrow\!-m$ and $q\!\rightarrow\!-q$ switching the two spinors inverting the sign of their mass term and charge: thus we recollect pairs of these fields together in terms of $\frac{1}{2}
(\mathbb{I}\!+\!\boldsymbol{\gamma})\psi^{N}\!=\!\psi_{R}^{N}$ and $\frac{1}{2}(\mathbb{I}
\!-\!\boldsymbol{\gamma})\psi^{N}\!=\!\psi_{L}^{N}$ together with the alternative $\frac{1}{2}
(\mathbb{I}\!+\!\boldsymbol{\gamma})\psi^{P}\!=\!\psi_{R}^{P}$ and $\frac{1}{2}(\mathbb{I}
\!-\!\boldsymbol{\gamma})\psi^{P}\!=\!\psi_{L}^{N}$ allowing us to go from the irreducible 
$2$-dimensional representation to the reducible $4$-dimensional representation; in this representation the discrete transformations above are also recollected as $\boldsymbol{\Gamma}
\!\rightarrow\!-\boldsymbol{\Gamma}$ and also $\psi\!\rightarrow\!\boldsymbol{M}\psi$ with 
$m\!\rightarrow\!-m$ and $q\!\rightarrow\!-q$ switching the two spinors inverting the sign of the mass term and charge: so we recollect the fields in terms of $\frac{1}{2}(\mathbb{I}\!+\! \boldsymbol{K})\psi\!=\!\psi^{N}$ and $\frac{1}{2}(\mathbb{I}\!-\!\boldsymbol{K})\psi\!=\!\psi^{P}$ allowing us to go from the $4$-dimensional representation up to the $8$-dimensional representation we will employ in this paper. Notice that these discrete symmetries, interchanging both mass and charge, are perfectly compatible with the heuristic definition of the matter/antimatter duality we decided to follow; this is of course not an accident, and it reflects the fact that within the general construction we have been building from the beginning the specific version of the matter/antimatter duality we will employ here fits perfectly. There is still the question about the possible problems that may appear by changing the sign of the mass beside that of the charge, but as we already said we will press on and we will see that this problem will be proven not to be a problem at all. In writing the conserved quantities, the property of the matrices to verify the Clifford algebra for $\frac{1}{2}$-spin spinor fields is a necessary and sufficient condition to get a completely antisymmetric spin density in its most general form as given by the following expression
\begin{eqnarray}
&S^{\rho\mu\nu}=\frac{i\hbar}{4}
\overline{\psi}\{\boldsymbol{\Gamma}^{\rho},\boldsymbol{\Sigma}^{\mu\nu}\}\psi
\label{spin}
\end{eqnarray}
which comes alongside to the non-symmetric energy density
\begin{eqnarray}
&T^{\mu}_{\phantom{\mu}\nu}=\frac{i\hbar}{2}
\left(\overline{\psi}\boldsymbol{\Gamma}^{\mu}\boldsymbol{D}_{\nu}\psi
-\boldsymbol{D}_{\nu}\overline{\psi}\boldsymbol{\Gamma}^{\mu}\psi\right)
\label{energy}
\end{eqnarray}
and also with the current given by
\begin{eqnarray}
&J^{\rho}=q\hbar\overline{\psi}\boldsymbol{K}\boldsymbol{\Gamma}^{\rho}\psi
\label{current}
\end{eqnarray}
where the spinor fields have to satisfy the matter field equations
\begin{eqnarray}
&i\hbar\boldsymbol{\Gamma}^{\mu}\boldsymbol{D}_{\mu}\psi-m\boldsymbol{K}\psi=0
\label{field}
\end{eqnarray}
and in which we recall that $q$ and $m$ are the charge and mass of the matter field and $\hbar$ is the Planck constant, and all the independent field equations are as many as all the independent degrees of freedom: the whole system of field equations up to the $8$-dimensional representation of the matter field equations is then given by the completely antisymmetric torsion-spin density field equations
\begin{eqnarray}
&Q^{\rho\mu\nu}=-a\frac{i\hbar}{4}
\overline{\psi}\{\boldsymbol{\Gamma}^{\rho},\boldsymbol{\Sigma}^{\mu\nu}\}\psi
\label{torsionspincoupling}
\end{eqnarray}
and the non-symmetric curvature-energy density field equations as 
\begin{eqnarray}
\nonumber
&\left(\frac{8\pi k}{a}\!-\!\frac{1}{2}\right)\!\left(\frac{1}{4}\delta^{\mu}_{\nu}Q^{2}
\!-\!\frac{1}{2}Q^{\mu\alpha\sigma}Q_{\nu\alpha\sigma}
\!+\!D_{\rho}Q^{\rho\mu}_{\phantom{\rho\mu}\nu}\right)
\!+\!\left(G^{\mu}_{\phantom{\mu}\nu}-\frac{1}{2}\delta^{\mu}_{\nu}G
-\lambda\delta^{\mu}_{\nu}\right)+\\
&+8\pi k\left(F^{\rho\mu}F_{\rho\nu}-\frac{1}{4}\delta^{\mu}_{\nu}F^{2}\right)
=8\pi k\frac{i\hbar}{2}
\left(\overline{\psi}\boldsymbol{\Gamma}^{\mu}\boldsymbol{D}_{\nu}\psi
-\boldsymbol{D}_{\nu}\overline{\psi}\boldsymbol{\Gamma}^{\mu}\psi\right)
\label{curvatureenergycoupling}
\end{eqnarray}
together with the gauge-current field equations
\begin{eqnarray}
&\frac{1}{2}F_{\alpha\mu}Q^{\alpha\mu\rho}+D_{\sigma}F^{\sigma\rho}
=q\hbar\overline{\psi}\boldsymbol{K}\boldsymbol{\Gamma}^{\rho}\psi
\label{gaugecurrentcoupling}
\end{eqnarray}
which are such that the conservation laws (\ref{conservationspin}-\ref{conservationenergy}) and (\ref{conservationcurrent}) are in fact verified whenever the matter field equations are given in the following form
\begin{eqnarray}
&i\hbar\boldsymbol{\Gamma}^{\mu}\boldsymbol{D}_{\mu}\psi-m\boldsymbol{K}\psi=0
\label{matterequations}
\end{eqnarray}
with $\lambda$ and $m$ seen as parameters and with four independent fields accounting for four different coupling constants $a$, $k$, $q$ and $\hbar$, and the independent field equations are as many as the different universal constants. Finally we notice that the $8$-dimensional representation is the one in which all possible degrees of freedom are written altogether, in what is the most compact way possible.
\subsection{Non-Linear Potentials for Physical Fields}
In the previous section we have obtained the dynamical field equations for the model we would like to study next, and we have seen that proceeding guided only by generality arguments the system of field equations turned out to be very comprehensive, accounting for both matter and antimatter as described by a single $8$-dimensional matter field: this situation should be regarded as natural, but it came at the cost of allowing negative mass terms into the matter field equations, which could create problems. Nevertheless, we will next show that the appearance of negative mass terms in the field equations does not imply that there will be negative mass states among their solutions whatsoever.

To proceed in our investigation, first we will decompose the system of field equations, and in order to do so we recall that since torsion is a tensor, all torsional quantities can be decomposed in terms of the corresponding torsionless quantities plus additional torsional contributions that will be converted by means of the algebraic torsion-spin field equations given by the expressions
\begin{eqnarray}
&Q^{\rho\mu\nu}\varepsilon_{\rho\mu\nu\alpha}
=\frac{3a}{2}\hbar\overline{\psi}\boldsymbol{\Gamma}_{\alpha}\boldsymbol{\Gamma}\psi
\label{torsion-spin}
\end{eqnarray}
into spinorial potentials in the gravitational field equations for the Ricci tensor
\begin{eqnarray}
\nonumber
&\left(R_{\mu\nu}+\lambda g_{\mu\nu}\right)
+8\pi k\left(F^{\rho}_{\phantom{\rho}\mu}F_{\rho\nu}-\frac{1}{4}g_{\mu\nu}F^{2}\right)
=-4\pi km\overline{\psi}\boldsymbol{K}\psi g_{\mu\nu}+\\
&+8\pi k\frac{i\hbar}{4}\left(\overline{\psi}\boldsymbol{\Gamma}_{\mu}\boldsymbol{\nabla}_{\nu}\psi
+\overline{\psi}\boldsymbol{\Gamma}_{\nu}\boldsymbol{\nabla}_{\mu}\psi
-\boldsymbol{\nabla}_{\nu}\overline{\psi}\boldsymbol{\Gamma}_{\mu}\psi
-\boldsymbol{\nabla}_{\mu}\overline{\psi}\boldsymbol{\Gamma}_{\nu}\psi\right)
\label{curvature-energy}
\end{eqnarray}
and the electrodynamic field equations as
\begin{eqnarray}
&\nabla_{\sigma}F^{\sigma\rho}=q\hbar\overline{\psi}\boldsymbol{K}\boldsymbol{\Gamma}^{\rho}\psi
\label{gauge-charge}
\end{eqnarray}
while the matter field equations are given instead by
\begin{eqnarray}
&i\hbar\boldsymbol{\Gamma}^{\mu}\boldsymbol{\nabla}_{\mu}\psi
+\frac{3a}{16}\hbar^{2}\overline{\psi}\boldsymbol{\Gamma}_{\mu}\boldsymbol{\Gamma}\psi
\boldsymbol{\Gamma}^{\mu}\boldsymbol{\Gamma}\psi-m\boldsymbol{K}\psi=0
\label{spinor}
\end{eqnarray}
showing that the gravitational and electrodynamic field equations are identical to those we would have in the torsionless case but the matter field equations are identical to those we would have if we were with no torsion but with spin-spin self-interactions of the matter field having the Nambu-Jona--Lasinio structure and an undetermined coupling constant given by the $a$ parameter; this system of field equations can also be decomposed from the $8$-dimensional form down to the $4$-dimensional form in either the chiral representation or the standard representation of the first type as $\overline{\psi}\!=\!(\overline{e},\overline{p})$ in terms of which we have that the gravitational field equations written for the Ricci tensor are given according to
\begin{eqnarray}
\nonumber
&\left(R_{\mu\nu}+\lambda g_{\mu\nu}\right)
+8\pi k\left(F^{\rho}_{\phantom{\rho}\mu}F_{\rho\nu}-\frac{1}{4}g_{\mu\nu}F^{2}\right)
=-4\pi k(m\overline{e}e\!-\!m\overline{p}p)g_{\mu\nu}+\\
\nonumber
&+8\pi k\frac{i\hbar}{4}\left(\overline{e}\boldsymbol{\gamma}_{\mu}\boldsymbol{\nabla}_{\nu}e
+\overline{e}\boldsymbol{\gamma}_{\nu}\boldsymbol{\nabla}_{\mu}e
-\boldsymbol{\nabla}_{\nu}\overline{e}\boldsymbol{\gamma}_{\mu}e
-\boldsymbol{\nabla}_{\mu}\overline{e}\boldsymbol{\gamma}_{\nu}e\right)+\\
&+8\pi k\frac{i\hbar}{4}\left(\overline{p}\boldsymbol{\gamma}_{\mu}\boldsymbol{\nabla}_{\nu}p
+\overline{p}\boldsymbol{\gamma}_{\nu}\boldsymbol{\nabla}_{\mu}p
-\boldsymbol{\nabla}_{\nu}\overline{p}\boldsymbol{\gamma}_{\mu}p
-\boldsymbol{\nabla}_{\mu}\overline{p}\boldsymbol{\gamma}_{\nu}p\right)
\label{gravitationa}
\end{eqnarray}
and the electrodynamic field equations as
\begin{eqnarray}
&\nabla_{\sigma}F^{\sigma\rho}=\hbar(q\overline{e}\boldsymbol{\gamma}^{\rho}e
\!-\!q\overline{p}\boldsymbol{\gamma}^{\rho}p)
\label{electrodynamics}
\end{eqnarray}
while the matter field equations are given instead by
\begin{eqnarray}
&i\hbar\boldsymbol{\gamma}^{\mu}\boldsymbol{\nabla}_{\mu}e
+\frac{3a}{16}\hbar^{2}\left(\overline{e}\boldsymbol{\gamma}_{\mu}\boldsymbol{\gamma}e
\!+\!\overline{p}\boldsymbol{\gamma}_{\mu}\boldsymbol{\gamma}p\right)
\boldsymbol{\gamma}^{\mu}\boldsymbol{\gamma}e-me=0
\label{matter}\\
&i\hbar\boldsymbol{\gamma}^{\mu}\boldsymbol{\nabla}_{\mu}p
+\frac{3a}{16}\hbar^{2}\left(\overline{e}\boldsymbol{\gamma}_{\mu}\boldsymbol{\gamma}e
\!+\!\overline{p}\boldsymbol{\gamma}_{\mu}\boldsymbol{\gamma}p\right)
\boldsymbol{\gamma}^{\mu}\boldsymbol{\gamma}p+mp=0
\label{antimatter}
\end{eqnarray}
in which we see that the mass term $m$ appears with opposite sign whenever we have either the matter or the antimatter field: this might produce negative mass contributions in the gravitational field equations. Nevertheless, we also have to take into account that anywhere the mass term is present with either a positive or a negative sign it is always accompanied by a bilinear spinor with a sign that is undefined, so that if it were $\overline{e}e\!\geqslant\!0$ and $\overline{p}p\!\leqslant\!0$ there will only be positive mass contributions in the gravitational field equations, and as a matter of fact, this happens whenever the relationship 
$p\!=\!\boldsymbol{\gamma}e$ holds; but such a constraints is precisely what we have if we want that the spinor field $e$ be the solution of the former matter field equation and that the spinor field $p$ be the solution of the latter matter field equation. This is not an accident, as it can be acknowledged by recalling that the former matter field equation is just the latter matter field equation after the discrete transformation we have discussed here above.

Nevertheless, to see that there are only positive mass contributions, or more in general positive energy contributions, as source of the gravitational field equations, it is better to consider the weak-gravitational low-speed limit given by the assumption $g_{tt}\!\approx\!1\!+\!2V$ in terms of which the gravitational field equations have only the time-time component given by the well known Newton law
\begin{eqnarray}
&\vec{\nabla}\!\cdot\!\vec{\nabla}V\!\approx\!-\lambda\!
+\!4\pi k\left[\vec{E}\!\cdot\!\vec{E}\!+\!\vec{B}\!\cdot\!\vec{B}
\!+\!m(u^{\dagger}u\!+\!v^{\dagger}v)\right]
\label{gravityweakslowspeed}
\end{eqnarray}
and for $F_{01}\!=\!E_{1}$, $F_{02}\!=\!E_{2}$, $F_{03}\!=\!E_{3}$ and $F_{32}\!=\!B_{1}$, $F_{13}\!=\!B_{2}$, $F_{21}\!=\!B_{3}$ the electromagnetic field equations have only the time component as the Gauss law
\begin{eqnarray}
&\vec{\nabla}\!\cdot\!\vec{E}\approx q\hbar(u^{\dagger}u\!-\!v^{\dagger}v)
\label{electrostatic}
\end{eqnarray}
where the spinors are given in standard representation for stationary configuration of energy $E$ with low-speed condition $(E^{2}\!-\!m^{2})\!\approx\!2m(E\!-\!m)$ and explicitly decomposed form given by $e^{\dagger}\!\approx\!(u^{\dagger},0)$ and $p^{\dagger}\!\approx\!(0,v^{\dagger})$ thus verifying the Schr\"{o}dinger field equations that are given according to the expressions
\begin{eqnarray}
\nonumber
&\frac{\hbar^{2}}{2m}\vec{\boldsymbol{\nabla}}\!\cdot\!\vec{\boldsymbol{\nabla}}u
\!+\!\frac{q\hbar^{2}}{2m}\vec{B}\!\cdot\!\vec{\boldsymbol{\sigma}}u
-\frac{9a^{2}\hbar^{4}}{512m}(u^{\dagger}\vec{\boldsymbol{\sigma}}u\!+\!v^{\dagger}\vec{\boldsymbol{\sigma}}v)\!\cdot\!
(u^{\dagger}\vec{\boldsymbol{\sigma}}u\!+\!v^{\dagger}\vec{\boldsymbol{\sigma}}v)u-\\
&-\frac{3a\hbar^{2}}{16}(u^{\dagger}\vec{\boldsymbol{\sigma}}u\!+\!v^{\dagger}\vec{\boldsymbol{\sigma}}v)\!\cdot\!\vec{\boldsymbol{\sigma}}u+(E\!-\!m)u\approx0
\label{material1}\\
\nonumber
&\frac{\hbar^{2}}{2m}\vec{\boldsymbol{\nabla}}\!\cdot\!\vec{\boldsymbol{\nabla}}v
\!-\!\frac{q\hbar^{2}}{2m}\vec{B}\!\cdot\!\vec{\boldsymbol{\sigma}}v
-\frac{9a^{2}\hbar^{4}}{512m}(u^{\dagger}\vec{\boldsymbol{\sigma}}u\!+\!v^{\dagger}\vec{\boldsymbol{\sigma}}v)\!\cdot\!
(u^{\dagger}\vec{\boldsymbol{\sigma}}u\!+\!v^{\dagger}\vec{\boldsymbol{\sigma}}v)v-\\
&-\frac{3a\hbar^{2}}{16}(u^{\dagger}\vec{\boldsymbol{\sigma}}u\!+\!v^{\dagger}\vec{\boldsymbol{\sigma}}v)\!\cdot\!\vec{\boldsymbol{\sigma}}v+(E\!-\!m)v\approx0
\label{material2}
\end{eqnarray}
as it is possible to check by following the usual calculations: notice that the low-speed limit gives only positive masses in both forms of the Schr\"{o}dinger matter field equations; then, after the macroscopic average is taken, we only find positive inertia in the Newton law, as expected. We remark that it is precisely the opposite sign of the mass term that determines the fact that within the spinor field the physical semi-spinorial components occupy different positions according to $e^{\dagger}\!\approx\!(u^{\dagger},0)$ and $p^{\dagger}\!\approx\!(0,v^{\dagger})$ when the low-speed limit is taken.

That the gravitational field equations have only positive energy so long as the spinors have low-speed approximation defined with only positive masses is rather intuitive, since the gravitational field equations are sourced by the energy density that in the low-speed limit becomes the gravitational mass and this is in turn proportional to the inertial mass, and so it is not surprising that the former is positive whenever the latter is positive: this fact has here a mathematical justification coming from the fact that, as already noticed, the positivity of the energy density is given by $\overline{e}e\!\geqslant\!0$ and $\overline{p}p\!\leqslant\!0$ which are verified whenever the two spinors are such that $p\!=\!\boldsymbol{\gamma}e$ while, as also already remarked, the positivity of the inertial mass comes from the fact that the low-speed approximation is given according to $e^{\dagger}\!\approx\!(u^{\dagger},0)$ and $p^{\dagger}
\!\approx\!(0,v^{\dagger})$ which take place as $e$ and $p$ are solutions corresponding to opposite signs of the mass term; therefore, the positivity of the energy density comes with that of the inertial mass since $p\!=\!\boldsymbol{\gamma}e$ holds precisely because $e$ and $p$ are solutions corresponding to an opposite-sing mass term in the matter field equations. We will deepen this discussion later on.
\subsection{Plane-Wave Solutions for Matter Fields}
In what we have done so far, we have seen that, although we started with a system of field equations in which the matter field equations had mass terms with both signs, at the end of the calculations only positive mass states were found to be present; this was best seen in the non-relativistic limit, where the two opposite-mass matter field equations gave rise to the same positive mass in non-relativistic matter field equations, the Schr\"{o}dinger equation: this shows that even if we started from both positive and negative mass terms eventually only positive mass states are found, and so no problem of bad mass states is actually present. This is clear from the equations, but it would be nicer if we could comprehend why this happens, and next we will see the reason for this.

To see that, consider plane-waves $i\hbar\boldsymbol{D}_{\mu}\psi\!=\!P_{\mu}\psi$ as usual; such a definition of plane-wave solutions comes directly from the plane-wave structure of quantum fields encoded in the fact that in the non-gravitational case these relationships can be split according to $i\hbar\partial_{t}\psi\!=\!E\psi$ and $i\hbar\vec{\nabla}\psi\!=\!-\vec{P}\psi$ which have to be valid in order to ensure that the commutation relationships between energy and time and between momenta and positions be respected: therefore such plane-wave structure is based on solid quantum grounds. When these are used to write the matter field equations in the momentum representation, we see that the matter field equations are $P_{\mu}\boldsymbol{\Gamma}^{\mu}\psi
\!-\!m\boldsymbol{K}\psi\!=\!0$ and because we are dealing with a massive field it is always possible to boost into its rest frame where these equations reduce to the form $(\boldsymbol{\Gamma}^{0}\!-\!\boldsymbol{K})\psi\!=\!0$ which decompose according to the two complementary constraints $(\boldsymbol{\gamma}^{0}\!-\!\mathbb{I})e\!=\!0$ and $(\boldsymbol{\gamma}^{0}\!+\!\mathbb{I})p\!=\!0$ giving in standard representation the structure $e^{\dagger}\!\approx\!(u^{\dagger},0)$ and $p^{\dagger}\!\approx\!(0,v^{\dagger})$ as we already know, and showing once again that according to the sign of the mass term the physical components occupy a different position inside the spinor: in the case of positive-mass term the physical semi-spinorial component is in the upper position while in the case of negative-mass term the physical semi-spinorial component is in the lower position of the spinorial field. Flipping the sign of the mass term makes the physical semi-spinorial components rearrange within the spinorial field itself.

The problem that appeared to arise for negative-mass terms as mentioned above is due to the common misconception that negative-mass terms must give negative-mass states necessarily, but this is an invalid argument since whichever is the sign of the mass term, the matter field always contains positive-mass as well as negative-mass states, and the reason for which we do not have negative-mass states is not that they are never included but only that when field equations are imposed they are always suppressed; when we change the sign of the mass term what happens is not that negative-mass states are introduced, because in general they are already there beside the positive-mass states, but only that there is a rearrangement of these components within the spinor, and so when the matter field equations are imposed, it will always be the negative-mass state the one that will be suppressed: the difference is that the one with positive-mass term suppresses the lower component while the one with negative mass term suppresses the upper component, but in any case the negative-mass states are always those that will be vanished. Thus only positive-mass states, those corresponding to physical components, will survive; these are the semi-spinor fields that verify the positive-mass Schr\"{o}dinger equation, and then only positive masses could possibly be consistently present. Therefore, we have only positive masses but for two complementary types of matter fields, and not the two opposite signs for the mass term of a single type of matter field.
\subsubsection{Four-Fold Multiple-States for Massive Particles}
Let us summarize the path we have followed: our approach moved from general considerations, in terms of which we had decided to apply the heuristic definition of antimatter as matter with all quantum numbers reversed wholly, reverting the sign of the mass, beside that of the charge: the generality of our geometrical settings allowed us to do that, and as a final result we got a system of field equations, in which those for the matter field were allowed to have both signs in the mass term; nevertheless only positive-energy densities were present for the gravitational field, and only positive-mass states were solutions of the matter field equations. This is rather intriguing, because the extension of the heuristic definition of matter/antimatter we have decided to consider provides a situation in which the negative-energy problem is never met, and the negative-mass problem that may have been present is solved by the structure of the matter field equations automatically; therefore, there are only positive masses but for two complementary matter/antimatter field equations, and not positive/negative masses for a single matter field equation. Additionally, these two matter/antimatter field equations are such that their matter/antimatter solutions are precisely those that correspond to the definition of matter/antimatter that is commonly assigned, as we are going to show. To do this we will have to compare the present and the usual definitions of matter and antimatter. 

In our approach the two complementary matter/antimatter solutions are given according to $\mathrm{exp}[-iP_{\mu}x^{\mu}] e_{0}$ and $\mathrm{exp}[-iP_{\mu}x^{\mu}]\boldsymbol{\gamma}e_{0}$ for matter and antimatter, respectively, in terms of the same constant spinor $e_{0}$ depending only on the momentum components, and for both solutions, their energy density is positive-defined by construction; in the common interpretation of matter/antimatter, matter is defined by a spinor field $e$ and antimatter by $\boldsymbol{\gamma}^{2}e^{\ast}$ accompanied by the additional assumption for which the spinor fields have to be Grassmann-valued in order to ensure the positivity of the energy, but after some algebra is performed, it is a text-book exercise to demonstrate that the plane-wave solutions are given by $\mathrm{exp}[-iP_{\mu}x^{\mu}]e_{0}$ and $\mathrm{exp}[iP_{\mu}x^{\mu}]\boldsymbol{\gamma}e_{0}$ for matter and antimatter, respectively, in terms of the constant spinor $e_{0}$ only \cite{p-s}: the single difference in the solutions of the two approaches is in the sign of the complex unitary phase of the antimatter field, but such a complex unitary factor leaves no trace when computing the spin-sum completeness relationships, leading to identical scattering amplitudes. This still holds also in the non-relativistic limit.

As a consequence of this fact, the definition of matter/antimatter fields presented here and the definition of matter/antimatter fields that is commonly given are equivalent, whenever scattering amplitudes are considered.

Although we retain that our definition of matter/antimatter fields is logically simpler than the commonly accepted one, it would be interesting to seek for a more pragmatic way to discriminate the two approaches: in our definition of matter/antimatter, the two fields are individual solutions of two complementary matter field equations differing for the sign of the mass term, so the matter and antimatter degrees of freedom belong to two different spinors; in the common definition of matter/antimatter, the two fields are the two different solutions of a single field equation with a fixed mass term, so matter and antimatter degrees of freedom are contained in one spinor. In the usual approach energies are positive only if Grassmann variables are introduced, which means that charge-conjugation converts a solution into another solution, and that is why matter and antimatter belong to the same spinor; in our approach, energies are already positive with no need for Grassmann variables, but then the charge-conjugation does no convert a solution into another solution due to the non-linearities of the matter field equations, and that is why matter and antimatter have to be contained in two different spinors. In our approach, the non-linearities produce a halving of the space of the solutions of the matter field equations that is compensated by a doubling of the matter field equations themselves, in order to still have four independent degrees of freedom. This has implications when the non-relativistic limit is considered: in the usual approach we would have a single Schr\"{o}dinger matter field equation, whose solution is a semi-spinor describing both matter and antimatter; in our approach we would have two Schr\"{o}dinger matter field equations, whose solutions are one semi-spinor for matter and the other for antimatter. Thus from a purely theoretical point of view, if the common definition of matter/antimatter is true then both matter and antimatter would already be present in non-relativistic regimes, whereas if our definition of matter/antimatter is true then matter and antimatter can only make sense in a genuinely relativistic context. More importantly: in the common approach, matter and antimatter are complex conjugate of one another, and in particular neutral fields have to be real wave functions, while in the present approach, matter and antimatter are independent, and even neutral fields are described by intrinsically complex wave functions. This fact has two consequences.

On the one hand, according to our definition of matter/antimatter, the passage from one type of matter to the complementary one is not achieved in terms of a discrete transformation, nor have we Grassmann-valued fields, so that it is impossible to define complex-conjugate Grassmann-valued fields such as Majorana fields, and so we should have none of the consequences of their presence, such as the neutrino-less double-beta decay \cite{Avignone:1991nk}: this fact makes our approach scientifically sound since detection of neutrino-less double-beta decay would falsify it, although no evidence for this decay is available yet \cite{v-p}; on the other hand, that we do not need the spinor to be an eigenstate of charge conjugation in order for it to be neutral means that neutral fields should merely be defined in terms of the simple condition $q\!=\!0$ without any restriction on the wave function and therefore on their massiveness, leading to the fact that according to our scheme, neutral massive fields do not violate any constraint and thus there is no reason why they could not exist: hence, in the picture we have presented, there naturally is the place for a dark matter candidate, without the need of invoking exotic superstructures \cite{Ellis:2000bb}. The fact that in our approach any field, whether charged or simply neutral, is described by a complex wave function allows interference patterns typical of quantum physics to occur, since it is precisely the unitary complex phase that provokes interference of quantum states, and thus we expect that some of the most relevant contributions of the intrinsically complex nature of the wave function for neutral fields can be shown in experiments involving neutron interferometers; in these experiments, intrinsic non-linearities have already been investigated \cite{Audretsch:1982ux}, and it will be interesting to deepen this study in light of the perspective presented here. However, the neutron is not an elementary particle, and therefore studying its interference by treating it as a fundamental field might turn out to give results of limited validity.

The fact that we do not know any neutral massive and yet fundamental field renders it difficult to see further ways in which our results can be compared to experimental observations, and we hope that more pragmatically-inclined colleagues might be able to provide further answers.
\section*{Conclusion}
In this paper, we have introduced a general system of field equations constructed on the torsional completion of gravity with electrodynamics for matter fields in $8$-dimensional representation of the spinorial structure: we have decided to provide a new definition of the matter/antimatter duality by asking antimatter to be matter with all quantum numbers reversed, reverting the sign of the mass, beside that of the charge, and we have found: that with our definition of matter/antimatter, all energies are always positive with no additional assumption on the algebraic nature of the fields involved; that the fact that the two matter field equations have mass terms with opposite sign does not mean that matter is defined by a spinor with positive mass while antimatter is defined by a spinor with negative mass, but only that matter and antimatter are defined as two complementary spinors both with positive masses; that these two complementary spinors are different from those that would be found in the common definition of matter/antimatter, but they give rise to the same dynamics in terms of cross-sections and decay rates. The real difference is in the fact that in the common interpretation matter and antimatter are complex-conjugated of one another, so that Majorana fields described in terms of real wave functions can be defined, while in our interpretation matter and antimatter are described by two independent fields, so that no Majorana field is possible, and leading to the prediction that neutrino-less double-beta decay would not exist; we have also acknowledged that in this construction there naturally is a place for neutral massive particles that may be candidate for dark matter. Final comments about possible applications in the case of interferometry have been given.

The present approach allows an interpretation of the matter/antimatter duality that is altogether logically simpler than the commonly accepted one, although the fact that we do not know neutral massive fundamental fields makes it difficult to see how the two approaches can in practice be discriminated at least for the present state of our knowledge.


\begin{thebibliography}{30}
\bibitem{m-t-w}
C.Misner, K.Thorne, J.A.Wheeler, 
\textit{Gravitation} (Freeman, 1973).
\bibitem{ha}
K.~Hayashi,
\textit{Phys. Lett.} \textbf{B65}, 437 (1976).
\bibitem{xy}
Xin Yu, 
\textit{Astrophysical and Space Science} 
\textbf{154}, 321 (1989).
\bibitem{a-l}
J.~Audretsch, C.L\"{a}mmerzahl,
\textit{Class. Quant. Grav.} \textbf{5}, 1285 (1988).
\bibitem{m-l}
A.Macias, C.L\"{a}mmerzahl,
\textit{J. Math. Phys.} \textbf{34}, 4540 (1993).
\bibitem{f/1a}
L.Fabbri,
in \textit{Annales de la Fondation de Broglie:\\ Special 
Issue on Torsion} (Ed. Dvoeglazov, Fondation de Broglie, 2007).
\bibitem{f/1b}
L.Fabbri,
in \textit{Contemporary Fundamental Physics:\\ Einstein 
and Hilbert} (Ed. Dvoeglazov, Nova Science, 2011).
\bibitem{h-h-k-n}
F.W.Hehl, P.Von Der Heyde, G.D.Kerlick, J.M.Nester,\\
\textit{Rev. Mod. Phys.} \textbf{48}, 393 (1976).
\bibitem{h-o}
F.W.Hehl, Yu.N.Obukhov, in \textit{Annales de la 
Fondation de Broglie:\\ Special Issue on Torsion} (Ed. 
Dvoeglazov, Fondation de Broglie, 2007).
\bibitem{Capozziello:2011et} 
S.~Capozziello, M.~De Laurentis, 
\textit{Phys. Rept.} \textbf{509}, 167 (2011).
\bibitem{f/2a}
L.Fabbri,
\textit{Annales Fond. Broglie} \textbf{33}, 365 (2008).
\bibitem{f/2b}
L.Fabbri,
\textit{Int. J. Theor. Phys.} \textbf{51}, 954 (2012).
\bibitem{f/3}
L.Fabbri,
\textit{Mod. Phys. Lett. A} \textbf{27}, 1250028 (2012).
\bibitem{Baekler:2011jt}
P.~Baekler, F.~W.~Hehl,
\textit{Class. Quant. Grav.} \textbf{28}, 215017 (2011).
\bibitem{k}
F.~A.~Kaempffer,
\textit{Gen. Rel. Grav.} \textbf{7}, 327 (1976).
\bibitem{Fabbri:2011kq} 
L.Fabbri,
\textit{Gen. Rel. Grav.} \textbf{45}, 1285 (2013).
\bibitem{p-s}
M.E.Peskin, D.V.Schr\"{o}der,
\textit{Quantum Field Theory} (Perseus, 1995).
\bibitem{Avignone:1991nk} 
F.T.Avignone,
\textit{AIP Conf. Proc.} \textbf{243}, 1106 (1992).
\bibitem{v-p}
P.Vogel, A. Piepke,
\textit{Phys.Lett.B} \textbf{667}, 1 (2008).
\bibitem{Ellis:2000bb} 
J.R.Ellis,
\textit{AIP Conf. Proc.} \textbf{562}, 9 (2001).
\bibitem{Audretsch:1982ux} 
J.Audretsch, C.L\"{a}mmerzahl,
\textit{J. Phys. A} \textbf{16}, 2457 (1983).
\end{thebibliography}
\end{document}